# Amplitude and Phase Noise of Magnons

Sergey Rumyantsev[1,2], Michael Balinskiy[1], Fariborz Kargar[1], Alexander Khitun[1] and Alexander A. Balandin[1]

[1]Nano-Device Laboratory (NDL), Department of Electrical and Computer Engineering, University of California, Riverside, California, USA

[2]Center for Terahertz Research and Applications (CENTERA), Institute of High Pressure Physics, Polish Academy of Sciences, Warsaw, Poland

*Abstract*— The low-frequency amplitude and phase noise spectra of magnetization waves, *i.e.* magnons, was measured in the yttrium iron garnet (YIG) waveguides. This type of noise, which originates from the fluctuations of the physical properties of the YIG crystals, has to be taken into account in the design of YIG-based RF generators and magnonic devices for data processing, sensing and imaging applications. It was found that the amplitude noise level of magnons depends strongly on the power level, increasing sharply at the on-set of nonlinear dissipation. The noise spectra of both the amplitude and phase noise have the Lorentzian shape with the characteristic frequencies below 100 Hz.

## I. INTRODUCTION

The majority of devices for information processing and sensing applications are based on the charge transfer in different media, *e.g.* semiconductors, metals, or vacuum. Recently, a completely different approach – termed magnonics – received significant attention. It is based on manipulation of the spin currents carried by the magnetization waves – magnons – in electrical insulators [1-7]. Spin currents in insulators avoid Ohmic losses and, therefore, Joule heating. A number of new devices based on magnon propagation have already been proposed and demonstrated for data processing, sensing and imaging applications [8-10]. The operation frequency of magnonic devices ranges from the low GHz to THz frequencies. The key material for these devices is yttrium iron garnet (YIG). Despite the strong interest to magnonic devices, their low-frequency noise characteristics remained largely unexplored [11].

YIG has also been used for filters and resonators, operating at frequencies up to 26 GHz [12-15]. Noise properties of RF generators based on YIG spheres and delay lines were studied in details [16-20]. The noise of a generator depends on many factors, and the generator scheme may include several noise sources. The YIG crystal is only one of them. However, to the best of our knowledge, the noise properties YIG material have not been rigorously addressed yet. We have recently reported on the low-frequency amplitude noise in YIG waveguides [11]. The low-frequency noise is a ubiquitous phenomenon, present in all kinds of electronic materials and devices [21-33]. The low-frequency noise in magnonic devices is also an important metric, which deserves much more attention. In this paper, we report the results of the measurements of the low-frequency noise of magnons in a YIG waveguide, focusing on the *phase* noise.

## II. EXPERIMENTAL DETAILS

The YIG-film of 9.6 μm thickness and 1.5 mm × 13.5 mm in dimensions was grown on the gadolinium gallium garnet (GGG, $Gd_3Ga_5O_{12}$) substrate by the liquid phase epitaxy. The Ti/Au antennae for spin wave (SW) excitation and detection were fabricated on the surface of YIG-film waveguide (see inset in Figure 1). The devices were placed in a magnetic field created by the permanent neodymium magnet. Depending on the orientation of the magnetic field, the spin waveguide structure supports either the magneto-static surface spin waves (MSSWs) or backward volume magneto-static spin waves (BVMSWs). The MSSW can propagate either on the top surface of the YIG waveguide (surface waves) or at the interface between the YIG waveguide and GGG substrate (interface waves). The strength of the magnetic field corresponded to the ferromagnetic resonance (FMR) frequency of about 5 GHz. Antenna 1 or 3 were used to excite spin waves and antenna 2 was used as a receiver (see Figure 1).

In order to confirm the generation and propagation of magnon current through the electrically insulating waveguide we measured the S-parameters of the waveguide as a function of frequency and magnetic field. The measured S – parameters were compared with known dispersion laws for BVMSW and MSSW. Good agreement with the theory confirmed the type of propagating magnons, and allowed for tuning the $f_p$-H space parameters for the magnon noise studies. These data also confirmed that the signal is not a result of direct electromagnetic coupling between antennas.

Propagating in the waveguide, magnon current acquires variations in the amplitude and phase due to the fluctuations of the physical properties of the YIG thin film. In order to measure these fluctuations, the commercial Schottky diode (33330B Keysight Technologies Inc.) was connected to the antenna 2 (see Figure 1). The DC detected signal from the diode was amplified by the low noise amplifier and analyzed by the FFT spectrum analyzer.

### III. RESULTS AND DISCUSSIONS

Fourier transform of the signal from the diode yields the spectrum of the amplitude fluctuations. The amplitude noise of the magnons propagating along the interface between YIG waveguide and GGG substrate was the highest, and the lowest was the noise of the volume magnons. The high noise level of the interface magnons was explained by the YIG/GGG interface roughness, resulting in stronger fluctuations of the material parameters that govern magnon current propagation. The dependence of noise on power for interface and surfaces magnons reveals one or more maxima. The positions of the noise peaks correspond to the change of the slope of $S_{21}$ dependence on the power.

The spectra of the amplitude fluctuations had the shape of the Lorentzian, $S_V \sim 1/(1+f^2/f_c^2)$ with the characteristic corner frequency $f_c <$ 100-1000Hz. In the time domain, the magnon noise revealed itself as a random telegraph signal (RTS) noise. Very small changes in the input power of ~0.1dB led to the significant changes in the RTS noise and it spectrum. RTS noise is well known in electronic devices and charge density materials and devices [34-38]. Observation of RTS noise in the large magnon waveguides can be explained by the individual discrete macro events which contribute to both the noise and magnon dissipation processes [11].

Fluctuations of the speed of the magnon wave lead to the fluctuations of the phase of the wave. In order to measure the phase noise, the delay line itself with the three antennas, as shown in Figure 1, was used as a phase detector. For this purpose, the output of the external generator was split and fed to antennas 1 and 3, correspondingly (see Figure 1). The line for the antenna 3 included an attenuator and a phase shifter. Using the attenuator, the power on the output antenna 2 was adjusted to be equal when powered separately in each of two configurations. The resulting power on antenna 3 was equal to ~2 dBm. These two signals are merged at the plane, which corresponds to the antenna 3 (see Figure 1). Due to the interference, when both antennas 1 and 3 are powered, the power on the output is a function of the phase difference of these two signals. This phase difference is defined by the distance between antennas 2 and 3 and by the external phase shifter. Fluctuations of the magnon wave speed contribute to the phase difference, and are converted to the voltage fluctuations detected at the antenna 2. The symbols in Figure 2 show the DC voltage on the detector, which is proportional to the output power at the antenna 2 as a function of the phase shift adjusted by the external phase shifter. The solid line is a fit with $A \times cos^2(\Psi/2)$ function ($A$ is a fitting parameter). The derivative, $R$, of this function is a conversion coefficient, which determines the spectral noise density of the voltage fluctuations: $S_v = S_\Psi \times R^2$ ($S_\Psi$ is the spectral noise density of the phase fluctuations).

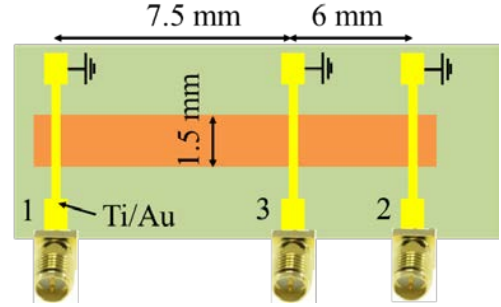

Fig.1. Schematic view of the YIG waveguide with three antennae.

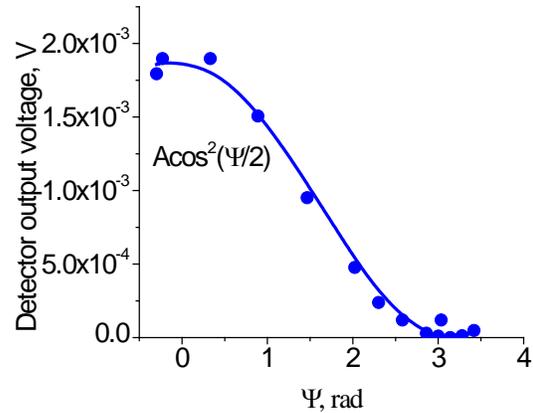

Fig. 2. DC voltage on the detector as a function of the phase shift adjusted by the phase shifter. The symbols show experimental data while the solid line is a fit with $A \times cos^2(\Psi/2)$ function.

Figure 3 shows the spectral noise density $S_v$ at $f$=10 Hz as a function of the phase difference. The blue symbols and line show the measured spectral nose density at $f$=10 Hz; the red symbols and dashed line show the level of the background noise. As one can see, the noise is minimal at the phase differences $\Psi \approx 0$ and $\Psi \approx \pi$. The noise at its maximum, i.e., at $\Psi \approx \pi/2$ is more that an order of magnitude higher. The conversion coefficient, $R$, has its maximum at this phase. Therefore, we can conclude that this amplitude noise is predominately due to the conversion of the phase noise.

Figure 4 shows the calculated phase noise $S_\Psi = S_v/R^2$. As seen, within the measurement accuracy, the phase noise is independent on $\Psi$, as expected. Although RTS noise was not found in the phase fluctuations, the noise spectra of the phase noise within the frequency range 10 Hz $< f <$ 1 kHz also had the form of the Lorentzian with the characteristic frequency within 10 Hz – 100 Hz (see Figure 5). With the excitation power of ~2 dBm on one of the antennas, the phase noise at the frequency of 10 Hz was measured to be around -68 dB/Hz.

We attributed the measured phase noise to the magnetization wave phase velocity fluctuations. The velocity fluctuations can be estimated as $S_v/V^2 = S_\Psi/\Psi_{int}^2$, where $\Psi_{int}$ is the phase difference gained by the spin wave between the antennas, which can be measured by the vector network analyzer. $\Psi_{int}$ was measured by method of substitution used "unwrapped phase" format of VNA. Total phase margin at operation frequency was first measured with our device, $\Psi_{DUT}$, and then with electrically short SMA connector which substituted our device, $\Psi_{conn}$. Interested value $\Psi_{int}$ was calculated as difference $\Psi_{DUT} - \Psi_{conn}$. We found $\Psi_{int}$=93 rad, which yields $S_v/V^2 \approx 2\times 10^{11}$ Hz$^{-1}$ at the frequency of the analysis $f$=10 Hz. This value of the velocity fluctuations can be used to estimate the phase fluctuations in the waveguides of an arbitrary length.

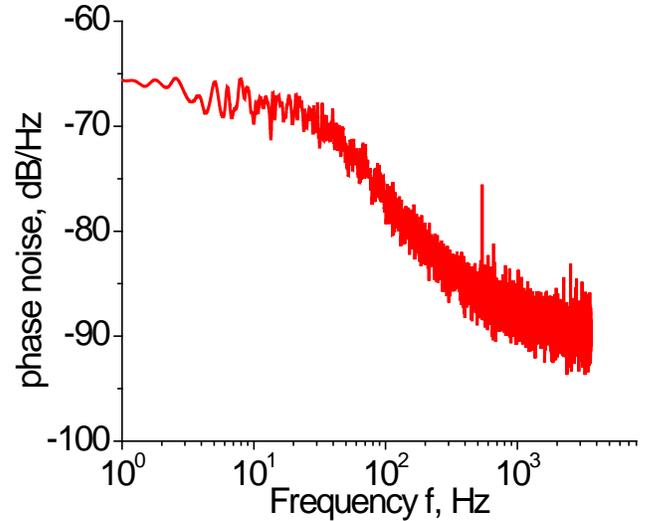

Fig. 5. Phase noise spectrum of magnons in YIG waveguide.

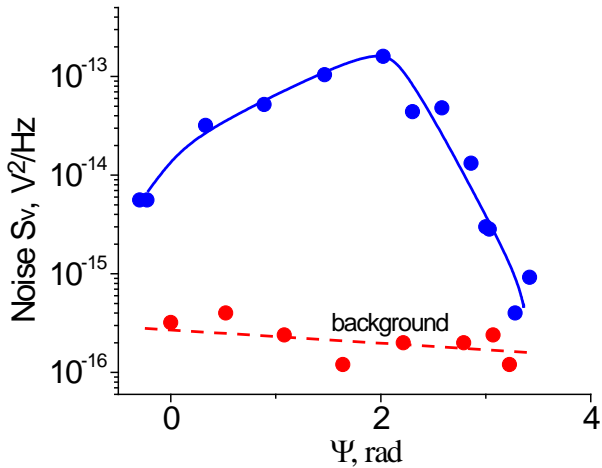

Fig. 3. Spectral noise density $S_v$ at $f$=10 Hz as a function of the phase difference. The dashed line and symbols show the background noise.

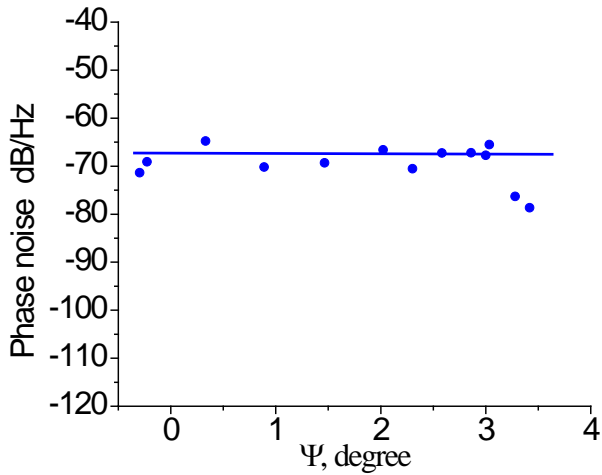

Fig. 4. Phase noise at $f$=10 Hz as a function of the phase shift.

## IV. CONCLUSIONS

In conclusion, the low-frequency noise of magnons, propagating as magneto-static surface spin waves, was measured in YIG waveguides at the frequencies $f$<1 kHz. The noise spectra had the Lorentzian shape with the characteristic frequencies below 100 Hz. At these frequencies, the noise of magnons sets the limit for data processing, sensing and imaging applications. It can also contribute to the phase noise of RF devices based on YIG crystals.

## ACKNOWLEDGMENTS

The work at UC Riverside was supported as part of the Spins and Heat in Nanoscale Electronic Systems (SHINES), and Energy Frontier Research Center funded by the U.S. Department of Energy, Office of Science, Basic Energy Sciences (BES) under Award No. SC0012670. S. R. acknowledges partial support from Center for Terahertz Research and Applications project carried out within the International Research Agendas program of the Foundation for Polish Science co-financed by the European Union under the European Regional Development Fund.